\documentclass[fleqn,usenatbib]{mnras}
\usepackage{newtxtext,newtxmath}
\usepackage[T1]{fontenc}
\usepackage{ae,aecompl}
\usepackage{graphicx}	% Including figure files
\usepackage{amsmath}	% Advanced maths commands
\usepackage{amssymb}	% Extra maths symbols
\usepackage{marvosym}
\usepackage{hyperref}
\usepackage{soul}
\newcommand{\beq}[1]{\begin{equation}\label{#1}}
\newcommand{\eeq}{\end{equation}}
\newcommand{\sub}[1]{_{\rm #1}}
\newcommand{\beqn}{\begin{eqnarray}}
\newcommand{\eeqn}{\end{eqnarray}}

\newcommand{\ar}{a\sub{r}}

\newcommand{\Rr}{R_\mathrm{Roche}}
\newcommand{\io}{i_\mathrm{o}}
\newcommand{\ith}{i_\mathrm{th}}
\newcommand{\emax}{e_\mathrm{max}}

\newcommand{\Rp}{R\sub{p}}

\newcommand{\fext}{f\sub{e}}
\newcommand{\fintt}{f\sub{i}}
\newcommand{\ARp}{A\sub{Rp}}

\newcommand{\Rpobs}{R\sub{p,obs}}

\newcommand{\hhl}[1]{\textcolor{black}{#1}}
\title[Anomalous light curves of young exorings]{Anomalous light curves of young tilted exorings}

\author[Sucerquia et al.]{
Mario Sucerquia\thanks{E-mail: \href{mailto:mario.sucerquia@udea.edu.co}{mario.sucerquia@udea.edu.co}}, J. A. Alvarado-Montes, Vanesa Ram\'irez  and Jorge I. Zuluaga\\
Solar, Earth and Planetary Physics Group (SEAP)\\ 
Instituto de F\'{\i}sica - FCEN, Universidad de Antioquia, Colombia\\ Calle 70 No. 52-21, Medell\'{\i}n, Colombia\\
}

\date{Accepted 2017 September 22. Received 2017 September 21; in original form 2017 July 11}

\pubyear{2017}

\begin{document}
\label{firstpage}
\pagerange{\pageref{firstpage}--\pageref{lastpage}}
\maketitle

% Abstract of the paper
\begin{abstract}
\hhl{Despite the success of discovering transiting exoplanets, several recently observed objects (e.g. KIC-8462852, J1407 and PDS-110) exhibit unconventional transit signals, whose appropriate interpretation in terms of a spherical single body has been challenging, if not impossible. In the aforementioned examples the presence of a ring-like structure has been proposed for explaining the unusual data. Thus, in this paper we delve into the dynamics of a tilted exoring disturbed by a third close companion, and the role that the Lidov-Kozai mechanism may have to explain irregular and anomalous transit signals of ringed planets, as well as the ring's early stages. To that end, we performed numerical simulations and semi-analytical calculations to assess the ring's dynamical and morphological properties, and their related transit observables. We found that tilted ringed structures undergo short-term changes in shape and orientation that are manifested as strong variations of transit depth and contact times, even between consecutive eclipses. Any detected anomaly in transit characteristics may lead to a miscalculation of the system's properties (planetary radius, semi-major axis, stellar density and others). Moreover, oscillating ring-like structures may account for the strangeness of some light-curve features in already known and future discovered exoplanets.}
\end{abstract}


%%%%%%%%%%%%%%%%%%%%%%%%%%%%%%%%%%%%%%%%%%%%%%%%%%
%%%%%%%%%%%%%%%%% BODY OF PAPER %%%%%%%%%%%%%%%%%%

\section{Introduction}
\label{sec:introduction}

Transit light-curve signals encode valuable information about the star, the exoplanet and its surroundings. An appropriate interpretation can provide us useful information to understand and characterize the distribution of planetary and stellar properties. Contrary to the straightforward explanation of simple lightcurves (the most frequent ones), in recent years we have discovered complex transit signals that challenge the well-established transit theories. However, given the complexity of these signals, new findings may emerge to reveal planetary processes that are absent in the Solar System.
 
\citet{Mamajek12} found a single transit \hhl{which} has been interpreted as a colossal ring around a sub-stellar object orbiting the star $J1407$. \hhl{Similarly, \cite{Osborn2017}, have proposed a ring-like structure as a feasible explanation for the two high-depth eclipses observed for PDS-110 system.} Recently, \citet{Boyajian16} reported an odd brightness stellar variability of \textit{KIC 8462852}, including dramatic and irregular dims without any analogous case in exoplanet observations. The latest object (that has been dubbed for short ``Tabby's star'') is the most intriguing case. Feasible explanations reported in literature, argue the possible detection of a younger star with coalescing material orbiting around it \citep*{lisse15}, a cloud of disintegrating comets \citep{Boyajian16,Bodman16} and a planetary debris field \citep{lisse15,Boyajian16}, as well as the combined effect of a huge planetary ring and co-orbital trojan asteroids \citep{Ballesteros2017mnras}. Beyond these well-known cases, there are lots of anomalous detections in photometric surveys \citep[and references therein]{Burke14}, that are awaiting for a proper confirmation and explanation.

The discovery of rings around exoplanets (hereafter exorings) is one of the expected outcomes of the forthcoming exoplanetary research. Although no exoring has been discovered so far (at least around a Jupiter-like planet), theoretical investigations \citep{Schlichting2011,Hedman2015} have showed that hypothetical `warm rings', i.e. rings made of refractory particles instead of volatile ones, could exist around the most abundant Jupiter-sized planets, namely close-in giants. The detection and characterization of a population of exorings around extrasolar giants is awaiting the development of novel and fast analysis techniques \citep{Zuluaga2015}, and more precise photometry provided by future missions such as \textit{JWST} and \textit{PLATO}.

Rings are complex structures and their transit signals will not be as simple as those of single planets. The most important dynamical mechanism affecting tilted rings is the so-called Lidov-Kozai mechanism (hereafter LKM) \citep{Lidov62,Kozai62}. LKM has been successfully applied to explain unconventional planetary system architectures, such as planets in highly eccentric orbits \citep{Takeda05}, formation of hot Jupiters \citep{Lithwick11}, among others. We hypothesize that LKM may be responsible of significant changes in the spatial configuration of exorings around close-in giant planets. These changes may occur in time-scales short enough to be detected in present and future lightcurves of ringed planets. Therefore, we propose that several anomalous transit signals could be explained as a consequence of the ring's complex dynamics.

This paper is organized as follows: Section \ref{sec:km} describes the behavior of rings under the LKM from an analytic approach. In Section \ref{sec:num-sim} we show the outcomes of numerical simulations, performed to asses the evolution of a tilted ring under the perturbing forces of the host star. Section \ref{sec:PR} discusses the effect of ring evolution on the  lightcurves. Finally, we devote Section \ref{sec:Discussion} to summarize and present the most remarkable conclusions of this work.
%%%%%%%%%%%%%%%%%%%%%%%%%%%%%%%%%%%%%%%%%%%%%%%%%%%%%%%
%Rings under the Kozai mechanism
%%%%%%%%%%%%%%%%%%%%%%%%%%%%%%%%%%%%%%%%%%%%%%%%%%%%%%%
\vspace{-0.4cm}
\section{The Lidov-Kozai mechanism}
\label{sec:km}
%KOZAI RINGS

%FFFFFFFFFFFFFFFFFFFFFFFFFFFFFFFFFFFFFFFFFFFFFFFFFFFFFFFFFFFFFFFFFFFFFF
%FIGURE: 
%FFFFFFFFFFFFFFFFFFFFFFFFFFFFFFFFFFFFFFFFFFFFFFFFFFFFFFFFFFFFFFFFFFFFFF
\begin{figure*}
\centering 
\includegraphics[scale=0.47]{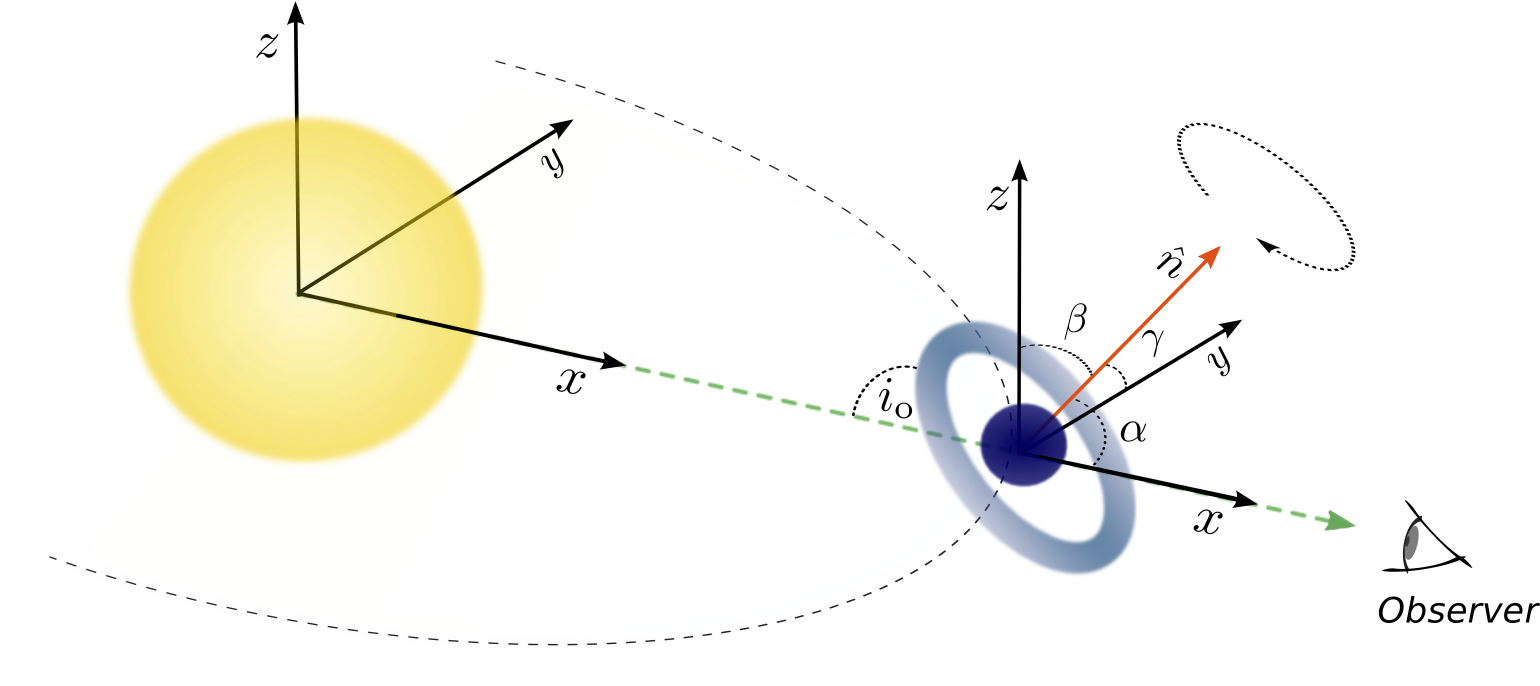}
\vspace{0.2cm}
\caption{Schematic representation of a ringed planet when passing in front of its host star. The initial inclination of the ring plane is $i_o$. The unitary vector $\hat{n}$ points in the direction of the average angular momentum of the ring particles. $\alpha$, $\beta$ and $\gamma$ are the angles between $\hat{n}$ and the coordinate axis $x$, $y$ and $z$, respectively. The $x$-axis points in the direction of the observer.}  
\label{fig:kozai-sch}
\end{figure*}
%FFFFFFFFFFFFFFFFFFFFFFFFFFFFFFFFFFFFFFFFFFFFFFFFFFFFFFFFFFFFFFFFFFFFFF

\begin{figure*}
\centering 
\includegraphics[scale=0.39]{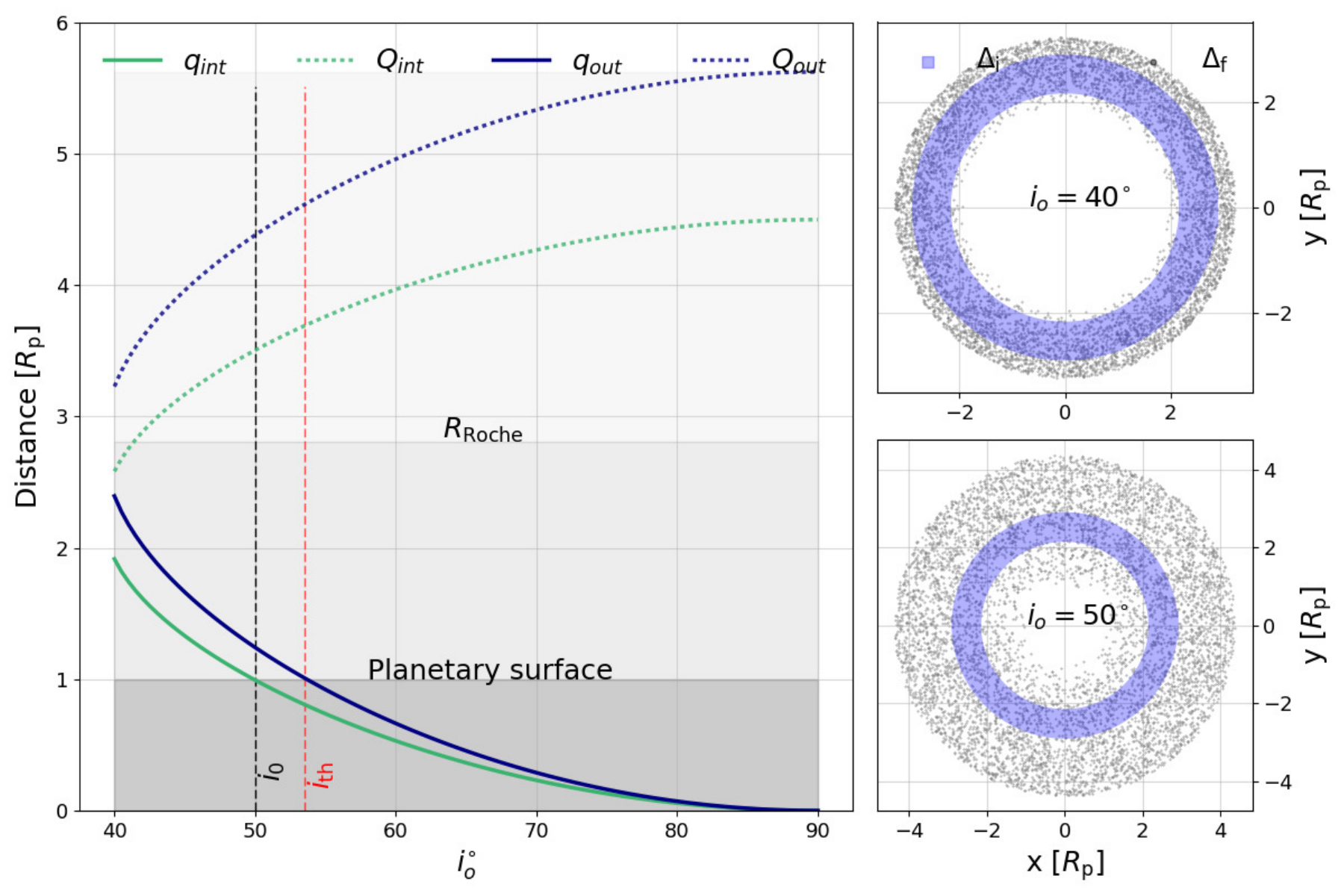}\\
\caption{\hhl{ {\it Left panel}: minimum periapsis $q$ and maximum apoapsis $Q$ reached by particles in the inner ({\it in} subscript) and outer edge ({\it out} subscript) of rings  with different initial inclinations $\io$, at $\emax$. Black and red vertical dotted lines represent the initial inclination  $\io$ used in our simulations and the threshold inclination $\ith$ (see \autoref{eq:th}) respectively. {\it Right panel}: original (blue strip) and estimated ring width (area with dots) for $\io = 40^{\circ}$, $50^\circ$ at $e_\mathrm{max}$}.
}  
\label{fig:iconstraints}
\vspace{0.2cm}
\end{figure*}

Particles in an initially tilted ring around a close-in planet are strongly affected by perturbations of the host star. When the initial inclination of the ring particles' orbits, as measured from the planet's orbit, is large enough (i.e. $\io\ga$39.2$^\circ$, \hhl{see a depiction of the system in \autoref{fig:kozai-sch}}), a cyclic exchange of angular momentum between the particles and the disturber arises. As a result, both the particle's orbital inclination and eccentricity undergo strong periodic oscillations. This is the Lidov-Kozai Mechanism. 
The maximum eccentricity reached by a particle $\emax$ depends only on $\io$ (see e.g. \citealt*{Ford00} and references therein):
\begin{eqnarray}
\emax &\simeq& \sqrt{1-(5/3) \cos^2{\io}}.
\label{eq:emax}
\end{eqnarray}
When particle orbits become more eccentric, the shape and ring's size $\Delta$ are also affected. Moreover, if $\io$ is large enough, particles in the inner part of the ring may collide with the planet and the ring is eroded over time. \hhl{The total depleting} of ring's particles is avoided only if $\io$ \hhl{is lower than} an inclination threshold $\ith$, which depends on planetary radius $\Rp$ and particle initial semi-major axis $\ar$ (see \autoref{fig:iconstraints}). The value of $\ith$ is computed from $\emax$, by imposing that the periapsis distance at maximum eccentricity $q_\mathrm{r}=\ar\;(1-\emax)$ be greater than $\Rp$, or equivalently:
\begin{eqnarray}
\cos^2 {\ith}&\simeq& \frac{3}{5} \left(\frac{\Rp}{\ar}\right)^2\left(2\;\frac{\ar}{\Rp}-1\right),
\label{eq:th}
\end{eqnarray}

\noindent here $\ar/\Rp$ is bounded between $\fintt$ and $\fext$, the inner and outer relative-radius of the ring, respectively. 
The initial value of $\fext$ is assumed smaller than or equal to the Roche radius $\Rr$.  However, when the outermost particles reach an eccentricity $\emax$, their apoapsis  $Q=\fext \, (1+\emax)$ become larger than that limit. Particles reaching those regions may coalesce to eventually drive the formation of moonlets \citep{Crida2012}. This process could also contribute to the loss of ring's particles.

The above mechanisms limit the range of initial sizes and inclinations that rings should have to survive LKM without loosing a significant fraction of their mass.  In \autoref{fig:iconstraints} we illustrate how the limits of those properties are determined and used for the numerical experiments described in the following section.
%FFFFFFFFFFFFFFFFFFFFFFFFFFFFFFFF
%FIGURE:  Allowed values
%FFFFFFFFFFFFFFFFFFFFFFFFFFFFFFFF
%FFFFFFFFFFFFFFFFFFFFFFFFFFFFFFf

%FFFFFFFFFFFFFFFFFFFFFFFFFFFFFFFFFFFFFFFFFFFFFFFFFFFFFFFFFFFFFFFFFFFFFF
%FIGURE: 
%FFFFFFFFFFFFFFFFFFFFFFFFFFFFFFFFFFFFFFFFFFFFFFFFFFFFFFFFFFFFFFFFFFFFFF
\begin{figure*}
{ 
\centering 
\includegraphics[scale=0.27]{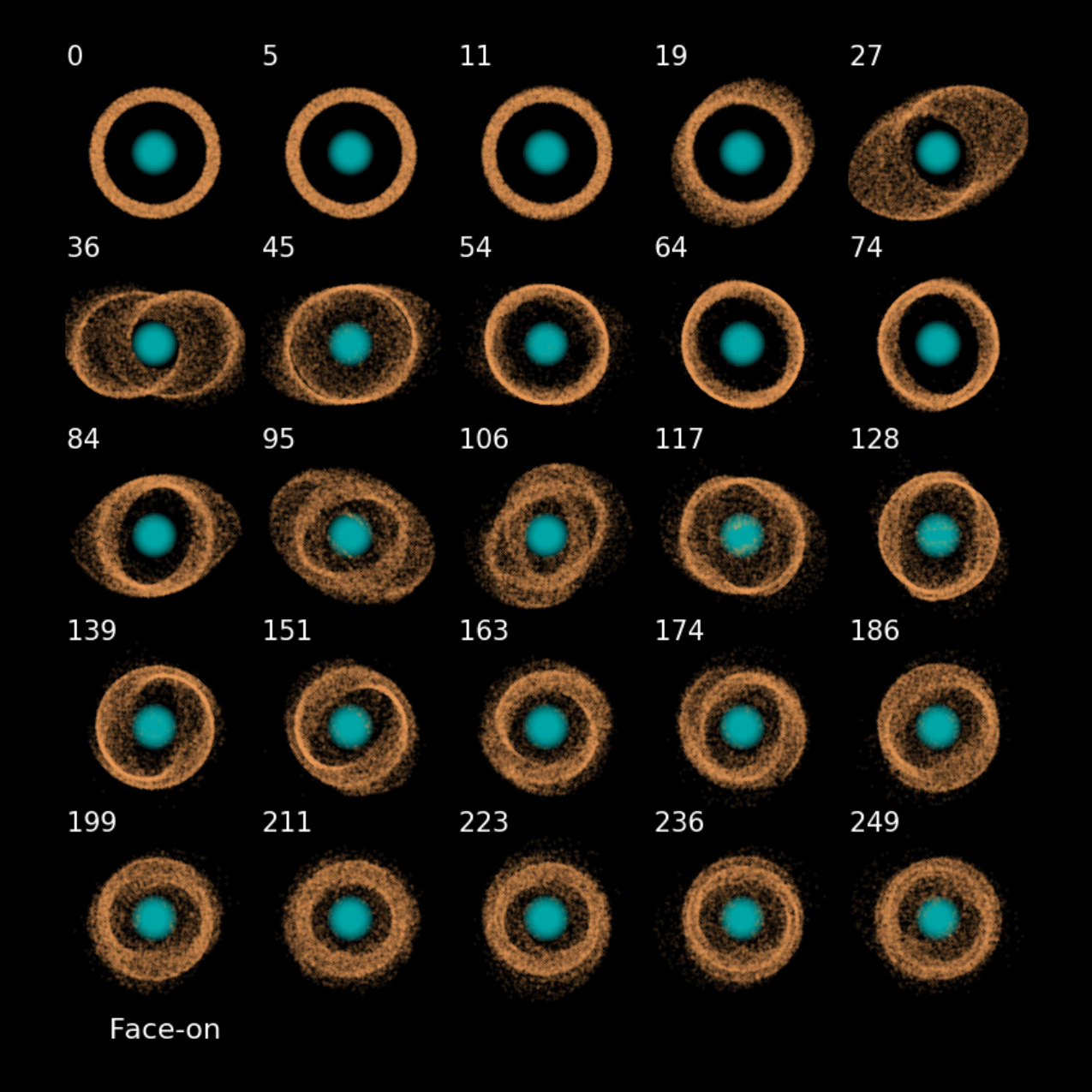}
\includegraphics[scale=0.27]{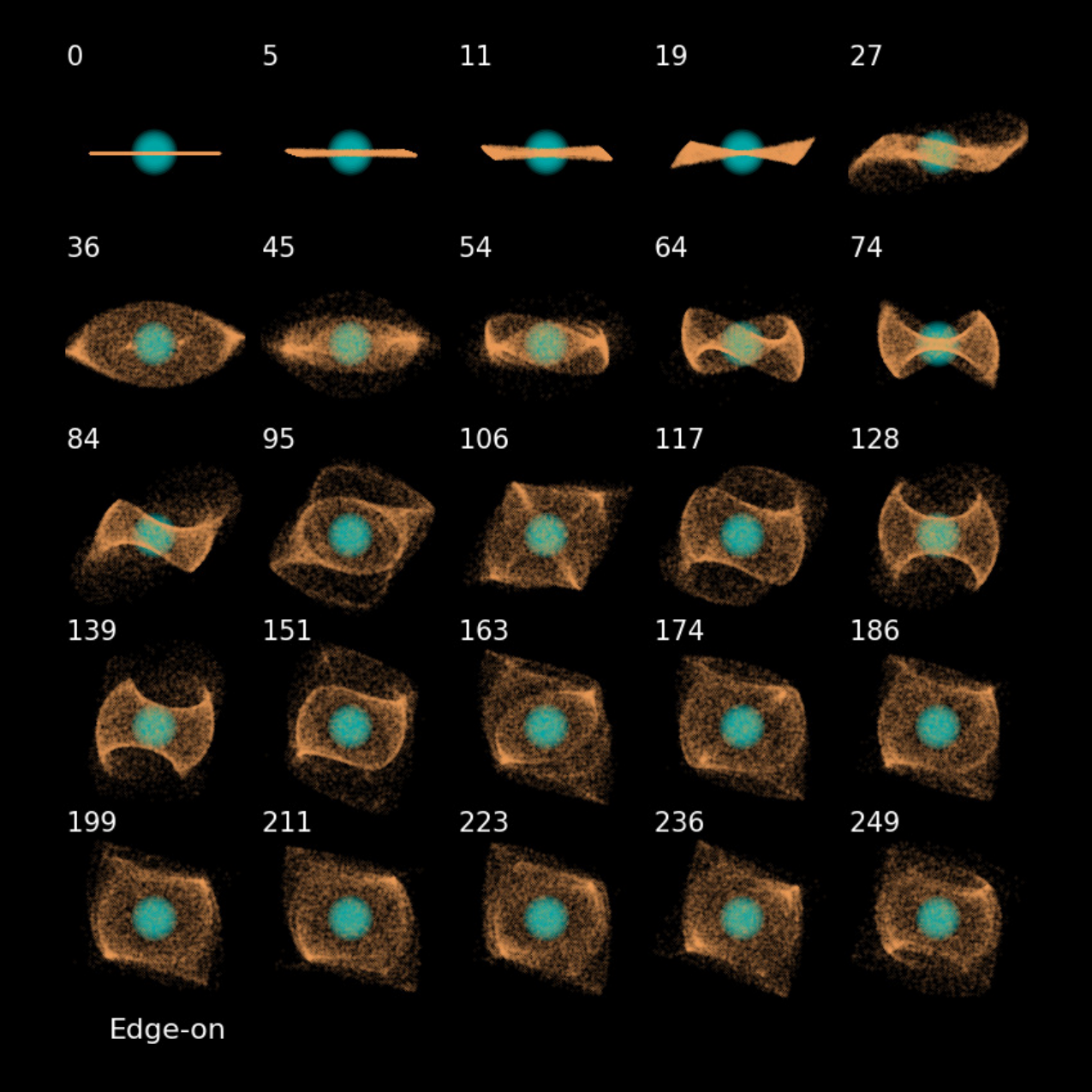}
\includegraphics[scale=0.27]{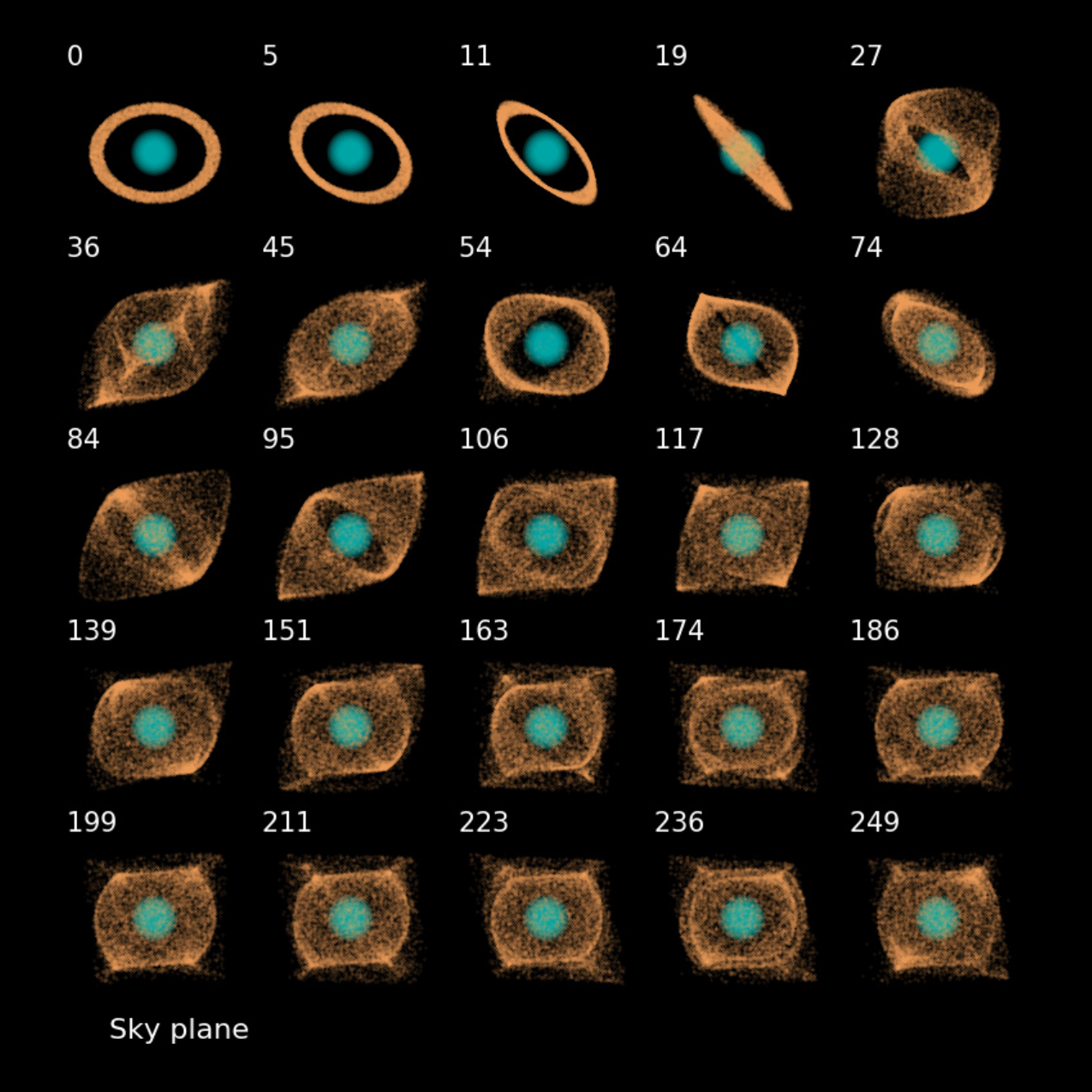}
\caption{\hhl{Snapshots of the evolution of a tilted ring for 250 transits. The ring's face-on, edge-on and on-sky projections are shown in the left, middle and right panel, respectively. The sizes are on scale.}}
\label{fig:confspace}
}
\end{figure*}
\vspace{-0.4cm}
%FFFFFFFFFFFFFFFFFFFFFFFFFFFFFFf
%%%%%%%%%%%%%%%%%%%%%%%%%%%%%%%%%%%%%%%%%%%%%%%%%%%%%%%
% NUMERICAL SIMULATIONS
%%%%%%%%%%%%%%%%%%%%%%%%%%%%%%%%%%%%%%%%%%%%%%%%%%%%%%%
\vspace{-0.4cm}
\section{Numerical Simulations}
\label{sec:num-sim}
To test our hypothesis we performed a set of numerical experiments using \href{https://janus.astro.umd.edu/HNBody}{\tt HNBODY} \citep{Rauch2012}, where \hhl {a collection of $10^4$ massless and collisionless particles}, initially distributed across a tilted ring, evolves under the gravitational forces of both a close-in Neptune-like planet ($M_\mathrm{p} = 10^{-4} \; M_{\odot}$, $\Rp = 2.5 \times 10^{4}\;\mathrm{km}$) and a Solar-mass star. The planet is located in a circular orbit with $a_\mathrm{p}=0.1$ au.

The ring has an initial inclination $\io=50^{\circ}$ and inner ($\fintt$) and outer ($\fext$) relative-radii of  2.11 and  2.81, respectively. The latter value corresponds to the Roche radius as computed for loose-bound bodies made of refractory material, i.e. densities about $2,000\,\mathrm{kg/m^3}$. 
\hhl{If we also assume a ring's total mass about $10^{19}$ kg (that of Saturn's rings), each particle will have a mass of about $10^{15}$ kg and a size of $\sim 5$ km.}
Particles' initial positions are randomly generated using a Poisson Disc Sampling algorithm \hhl{to avoid oversampled regions} (see e.g. \citealt{Cook1986} and references therein), and the system was allowed to evolve during 300 planetary orbital periods.

\hhl{Since our goal here is to estimate the time-scales and other broader effects of a massive oscillating disk in the light curve of a transiting planet, we have neglected the effect that self-gravitation, collisional damping and viscosity have on the ring's evolution (see eg. \citealt{Hyodo2017} and references there in). A more accurate approach should take into account these factors. Also, self-gravitation and collisional damping will act on reducing the dispersion of particles and circularizing their orbits.  Viscosity (which arises from translational, collisional and gravitational interchange of angular momentum among ring particles) will contribute to spread out the disk, increasing the mass-loss rate, and consequently reducing the ring's optical depth and damping its oscillations \citep{Daisaka2001}. }

\hhl{As a side note, it is worth to notice that including the gravitational interaction among ring particles will also drive the formation of ringlets and wakes. However, since the dynamical time-scales of these processes (several years) is of the order of ring oscillation damping, their role on the light-curve evolution could be negligible.  On the other hand, the interaction among ring particles could contribute to self-organize the oscillating ring. This could actually enhance the ring's signal, but also modify the oscillation's time-scales predicted here.}

A graphical illustration of the results of our numerical experiments are presented in \autoref{fig:confspace}. Spatial configuration of the \hhl{resulting structure} as viewed from different \hhl{projections} at the time of planetary transits are depicted in 25 snapshots (250 planetary periods). 
As it is evident in the \hhl{face-on} projection (left panel), the ring's width changes significantly whilst the whole structure oscillates. From a cursory comparison \hhl{between the first and last snapshot, we realize that $\Delta$ `expands' by a factor of $\sim$3, which is noticeable in the left panel (lower plot) of \autoref{fig:flux}, that gives a better insight about the broadness evolution.}
To follow the changes in the ring's orientation, we computed the ringlets' total angular momentum.  This quantity gives us the direction of the plane vector $\hat{n}$ (see \autoref{fig:kozai-sch}). The orientation of the ring (plotted in the upper panel of \autoref{fig:flux}) is calculated for the entire lifespan of the system, by means of the director cosines or director angles $\alpha$, $\beta$ and $\gamma$ \hhl{also depicted in the left panel (upper plot) of \autoref{fig:kozai-sch}.}
\vspace{-0.6cm}
%%%%%%%%%%%%%%%%%%%%%%%%%%%%%%%%%%%%%%%%%%%%%%%%%%%%%%%
%The effect in the lightcurves 
%%%%%%%%%%%%%%%%%%%%%%%%%%%%%%%%%%%%%%%%%%%%%%%%%%%%%%%
%SSSSSSSSSSSSSSSSSSSSSSSSSSSSSSSSSSSSSSSSSSSSSSS
\section{Effect in lightcurves}
\label{sec:PR}
\begin{figure*}
{ 
%\centering 
%
\includegraphics[scale=0.33]{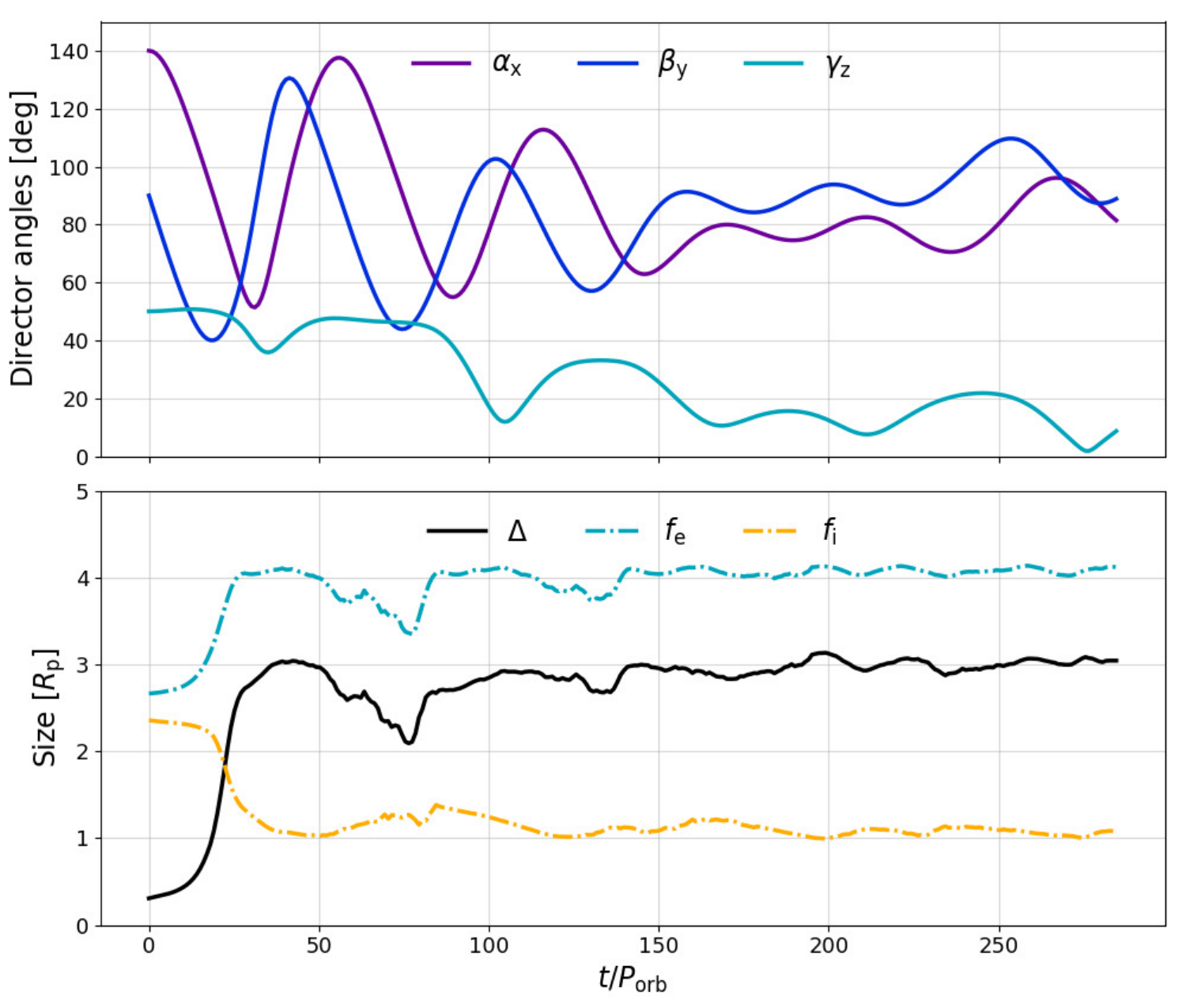}\hfill
\includegraphics[scale=0.33]{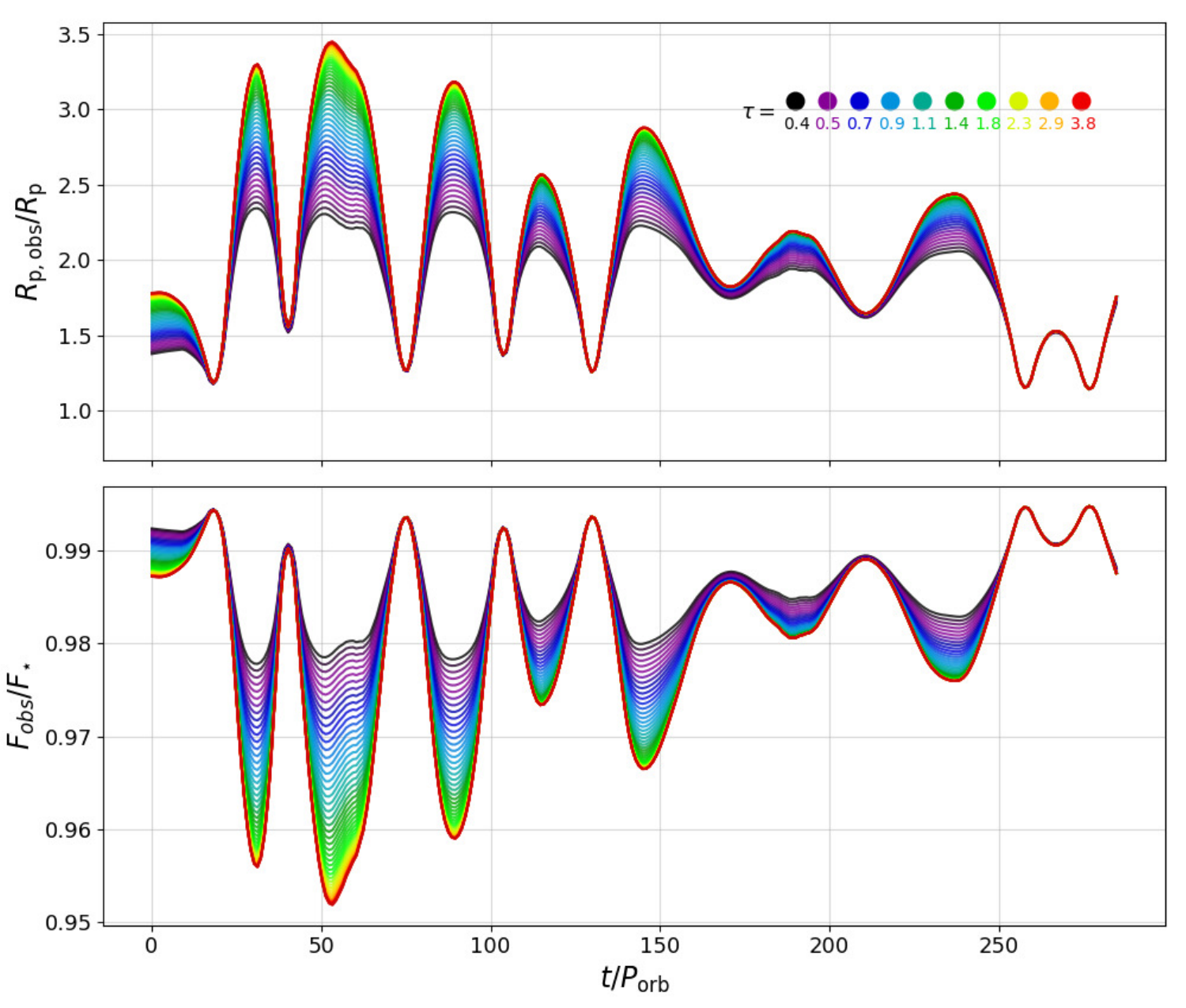}
\caption{{\textit{Left panel}}: upper plot,  evolution of the director angles ($\alpha$, $\beta$ and $\gamma$) of the total angular momentum of ring particles, $\hat{n}$; lower plot shows the evolution of the ring effective extent, $\fintt$ (yellow) and $\fext$ (green), and width (black solid line). {\textit{Right panel}}: upper plot, effective planet size as obtained from eclipse maximum depth, and minimum stellar flux (lower plot) as computed assuming different effective optical depths $\tau$.}
\label{fig:flux} 
}
\end{figure*}

%SSSSSSSSSSSSSSSSSSSSSSSSSSSSSSSSSSSSSSSSSSSSSSS
Numerical simulations suggest that a tilted ring subject to LKM, evolves towards a toroidal-like structure.  Computing the transit signal of such a complex body is not trivial. Furthermore, we have noticed in our simulations that in spite of the three dimensional dispersion, the vertical density of particles is strongly peaked around the instantaneous Laplace plane of the ring. As a first approximation we will model the transit signal of the oscillating rings as that of a unique disc-like structure, with internal and external radii $\fintt$, $\fext$, width $\Delta$ and projected inclination $i_{\sub R}=90^\circ-\alpha$ \hhl{(see the evolution of these parameters in the left panel of  \autoref{fig:flux})}. In order to take into account the variation in vertical dispersion, we computed the expected transit depth assuming different values of the effective opacity $\tau$.

To estimate the transit's observables, namely the observed minimum flux $F_\mathrm{obs}$, the transit depth $\delta=\ARp/A_\star$ (with $\ARp$ and $A_\star$ are the effective projected ring-planet and stellar area), and the equivalent planetary radius $\Rpobs \simeq \sqrt\delta$, we use Eqs. (1), (2) and (3) in \cite{Zuluaga2015}.

The results are presented in right panel of \autoref{fig:flux}.  As expected, significant oscillations arise both in  $F_\mathrm{obs}$ and $\Rpobs$, especially in the case of dense rings (higher values of $\tau$). In that case the transit of the oscillating rings is manifested by a maximum stellar dimming of $\sim5$ per cent, \hhl{corresponding to a planet's observed radius $\sim3.5$ times larger than a planet without rings}. It is worth to notice that the effective planetary radius estimated from the lightcurve, is consistent with those values predicted in the analytical results of \autoref{fig:iconstraints}, giving some confidence on the adopted approximation to compute the transit's observables. 

The transit signal is mainly determined by the evolution of the projected inclination $\alpha$. However, variations in the spatial extent and width of the ring create additional dips and peaks in the lightcurve. It is important to stress that the curves in the rightmost column of \autoref{fig:flux} only represent the flux and effective radius at the eclipse's centre, and not actual lightcurves. Moreover, the exact value of those maxima depends on the way that the optical depth changes in an oscillating ring.

Transit contact times also depend on the evolution of $\Delta$ and $\alpha$ (which is noticeable for different ring's projections shown in \autoref{fig:confspace}).  In a real case these variations give rise to uncertainties in the planetary orbital period, as well as in the stellar density \cite{Zuluaga2015}.
%FIGURE: 
%FFFFFFFFFFFFFFFFFFFFFFFFFFFFFFFFFFFFFFFFFFFFFFFFFFFFFFFFFFFFFFFFFFFFFF
%FFFFFFFFFFFFFFFFFFFFFFFFFFFFFFFFFFFFFFFFFFFFFFFFFFFFFFFFFFFFFFFFFFFFFF

%%%%%%%%%%%%%%%%%%%%%%%%%%%%%%%%%%%%%%%%%%%%%%%%%%%%%%%
%Discussion 
%%%%%%%%%%%%%%%%%%%%%%%%%%%%%%%%%%%%%%%%%%%%%%%%%%%%%%%
%SSSSSSSSSSSSSSSSSSSSSSSSSSSSSSSSSSSSS
\vspace{-0.4cm}
\section{Summary and Discussion}
\label{sec:Discussion}

%SSSSSSSSSSSSSSSSSSSSSSSSSSSSSSSSSSSSS
The recent discovery of anomalous stellar lightcurves, have inspired the search for astrophysical systems and mechanisms which can drive non-trivial transit signals.  In some cases it has become common to appeal to transient phenomena without any parallel in the Solar System.  In this work we studied a dynamical process, namely the Lidov-Kozai mechanism (LKM), capable of producing a quasi-periodic alteration of the transit's observable features of a young ringed close-in planet.

For a given set of ring's initial properties ($\io$, $\fintt$ and $\fext$) LKM determines the evolution and final fate of a ring through three mechanisms: 1) depletion of material of the ring through collision of particles with the planet at their periapsides, 2) drifting of ring's particles beyond the Roche limit, potentially driving the coalescence of material and the formation of moonlets, and 3) alteration of the spatial distribution and thereby ring's geometrical parameters. Any of these mechanisms will be quite notorious in the corresponding planetary lightcurves.

LKM and its light-curve signatures are transient in nature. The light-curve depth substantially depends on the ring projected inclination $\alpha$, which as shown in the left panel (upper plot) of \autoref{fig:flux} tends to damp (driven by the damping of the physical ring inclination $\beta$).  In our case the time required for the amplitude of $\beta$ to decay to about 40 per cent ($t_\mathrm{40}$) is only $\sim 9 $ yr.  Therefore, the probability of observing a system undergoing this process is very small. However, as LKM oscillation period strongly depends on the distance to the perturbing object (e.g. the host star), the time-scales for outermost planets could be relatively large.  For instance, if we assume the same planetary mass but use distances of 0.2, 0.3 and 0.5 au,  $t_\mathrm{40}$ results in $\sim$ 20, 35 and 85 yr, respectively, increasing the chances to detect the predicted anomalies.

LKM could also explain the absence of any ring signature among the already close-in planets surveyed by \citet*{Heising2015} and \citet{Aizawa2017AJ}. Depending on the ring's initial mass, LKM may entirely obliterate an initial light ring.  On the other hand, if rings are massive enough, the induced eccentric orbits enhances the   collisional rate of particles, which may contribute to circularize and make more coplanar the particles' orbits (see discussion in section 5.4 by \citealt{Hyodo2017}). The latter mechanism, coupled with the mass-loss processes described before, may lead to final low-inclined and less-massive rings as those observed in the Solar System.

The results presented in this work may also shed light on the possible fate (and observational signatures) of irregular moons wiped out by tides (i.e. within the Roche limit). As shown here, the observation of anomalous lightcurves of close-in massive planets could be an evidence of these kinds of processes.  Therefore, including the effect of LKM when preparing present and future searches for exomoons and exorings around close-in transiting planets, could be an interesting task to pursue, if not mandatory. 
To notice here, our results are restricted to close-in planets.  However, LKM may also operate if the external perturber, instead of the host star, is a massive moon or even a close planetary companion.  In those cases the anomalous transit signals described here could also be observed in ringed planets located at considerably larger distances from their stars.

\hhl{Our simulations assume a dense ring without any substructure (ringlets, wakes or gaps). For instance, gaps could produce remarkable signatures in the lightcurve (as observed for the cases of J1407b and PDS-110), but LKM oscillations tends to erase any trace of them due to the significant induced changes in the orbits' eccentricity. That said, it could be interesting to study the effect that LKM has on the evolution of any ring substructure, but they have much larger formation time-scales than the processes considered here \citep{Daisaka2001}.  The gaps' formation could require the presence of moons that accrete or disperse the material in time-scales of many orbital periods. The ringlets' formation will require several diffusion time-scales \citep{Tremaine2003}. In this sense, our results are restricted to young discs where some of these effects are just happening or about to arise. Nevertheless, our outcomes (\autoref{fig:confspace}) exhibit some void and overdensed regions, which might be confounded in lightcurves with the aforementioned substructures, although these features do not remain over time.}

Finally, in the light of our results, it could also be possible to speculate about the nature of the hypothetical object orbiting {\it KIC-8462852}. Provided enough information one may fit the observable signatures to a \hhl{large} oscillating disc made of circumplanetary debris subject to LKM. We speculate that if consecutive observations evidence some signatures of \hhl{damped oscillations} in the observed transit depth, we could be witnessing for the first time the disruption of a moon and the birth of a new ringed exoplanet. However, the confirmation or disprove of this hypothesis is well beyond the scope of this paper.
\vspace{-0.4cm}
\section*{Acknowledgements} 
M.S. is supported by Doctoral Program of Colciencias and the CODI/UdeA. J.I.Z. is supported by Vicerrector\'ia de Docencia U de A. To Nadia Silva for the proofreading.
\vspace{-0.4cm}
% * <mario.sucerquia@udea.edu.co> 2017-07-06T00:52:09.589Z:
%
% ^.
%%%%%%%%%%%%%%%%%%%%%%%%%%%%%%%%%%%%%%%%%%%%%%%%%%
%%%%%%%%%%%%%%%%%%%% REFERENCES %%%%%%%%%%%%%%%%%%
% The best way to enter references is to use BibTeX:
\nocite{*}
\input{Sucerquia2017_MNRAS_LCR_v2.bbl}
% \bibliographystyle{mnras}
% \bibliography{references} % if your bibtex file is called example.bib
%%%%%%%%%%%%%%%%%%%%%%%%%%%%%%%%%%%%%%%%%%%%%%%%%%

% Don't change these lines
\bsp	% typesetting comment
\label{lastpage}
\end{document}